# Emerging Technologies in Mass Spectrometry Imaging

Julia H. Jungmann and Ron M.A. Heeren*

*FOM-Institute AMOLF, Science Park 104, 1098 XG Amsterdam, The Netherlands*

\* Address reprint requests to Ron M.A. Heeren, FOM-Institute AMOLF, Science Park 104, 1098 XG Amsterdam, The Netherlands, Tel: +31-20-7547 161, Fax: +31-20-7547 290, E-Mail: heeren@amolf.nl

**Abstract**

Mass spectrometry imaging (MSI) as an analytical tool for bio-molecular and bio-medical research targets, accurate compound localization and identification. In terms of dedicated instrumentation, this translates into the demand for more detail in the image dimension (spatial resolution) and in the spectral dimension (mass resolution and accuracy), preferably combined in one instrument. At the same time, large area biological tissue samples require fast acquisition schemes, instrument automation and a robust data infrastructure. This review discusses the analytical capabilities of an "ideal" MSI instrument for bio-molecular and bio-medical molecular imaging. The analytical attributes of such an ideal system are contrasted with technological and methodological challenges in MSI. In particular, innovative instrumentation for high spatial resolution imaging in combination with high sample throughput is discussed. Detector technology that targets various shortcomings of conventional imaging detector systems is highlighted. The benefits of accurate mass analysis, high mass resolving power, additional separation strategies and multimodal three-dimensional data reconstruction algorithms are discussed to provide the reader with an insight in the current technological advances and the potential of MSI for bio-medical research.

## (1) Introduction

Mass spectrometry imaging (MSI) measurements target the visualization of the spatial organization and the identification of molecular masses from bio-molecular surfaces [1-3]. Ideally, a single MSI experiment on one instrument would return an accurate bio-chemical map of a sample surface. It would identify and localize all compounds with a sub-cellular lateral resolution, a high mass accuracy and zepto-molar sensitivity. The analysis speed would be so high that large-area imaging and high sample throughput would be guaranteed. Not solely mass analysis would be performed over a wide mass range but also structural information on sample molecules would be revealed by on-the-fly data dependent tandem mass spectrometry. The localization of specific analytes of bio-molecular or bio-medical samples would be placed in the context of tissue and/or organ functionality using clever three-dimensional reconstruction algorithms.

Unfortunately, the ideal MSI experiment is far from reality and a single instrument unifying the above capabilities has not been realized to date. A variety of different types of imaging mass spectrometers deliver a range of specific analytical capabilities that can be used in concert. In these spectrometers, a clever piece of instrumentation or artificial intelligence enable measurements that approach one or few of the desirable attributes of the "ideal" MSI experiment. In addition, the field is pushing novel instrumental innovations and the integration of several analytical capabilities into one instrument. In addition novel multi-modal approaches are pursued that analyze one sample with different molecular imaging instruments. Subsequent data correlation analysis is used to combine the generated complementary information and to employ it for revealing more molecular detail.

This article briefly describes the current state of MSI. It identifies technological and methodological challenges and recent advances in mass spectrometry imaging technology. Emphasis is put on innovative instrumentation for high spatial resolution imaging in conjunction with high sample throughput. Detector technology that targets various shortcomings of conventional imaging detector systems is introduced. Accurate mass identification, enhanced analytical MSI capabilities and multimodal 3D data reconstruction algorithms are described and discussed to provide the reader with an insight in the current technological advances in mass spectrometry imaging.

**(2) MSI and its challenges in a nutshell.**

Mass spectrometry identifies compounds based on the atomic composition of the sample molecules and their charge state. Therefore, no detailed prior knowledge of the sample composition is required and analysis becomes feasible. The chemical identification is not limited by analyte pre-selection as, for instance, in imaging techniques based on fluorescent or radioactive labeling. Hence, the technique itself does not introduce any functional changes on the bio-molecules under investigation. The combination of unlabelled identification and analyte localization within a sample provides the possibility to visualize and understand bio-molecular modifications and pathways. Depending on the mass analyzer, the detection range of this technique covers single atoms up to macromolecules. In principle, MSI can detect and identify hundreds of compounds with sub-micrometer resolution from complex biological samples surfaces, while maintaining a relatively high throughput. These features of MSI provide a solid basis to molecular pathology. Currently, MSI is applied to the fields of proteomics [4, 5], lipidomics [6-8] and metabolomics [9-11]. In addition, disease studies like the fundamental understanding of the bio-chemistry of neurodegenerative diseases [12, 13] or cancer [14], drug distribution studies [11, 15] and forensics [16, 17], among others, benefit from the information revealed by MSI.

Compound localization and identification form the core of MSI as an analytical tool for biomedical research. In terms of MSI instrumentation, these requirements translate into the demand for more detail in the image dimension (spatial resolution) and in the spectral dimension (mass resolution, accuracy and tandem MS capabilities for compound identification) –preferably combined in a single instrument. At the same time, large area tissue samples call for fast acquisition schemes, instrument automatization and robust data infrastructure, among other desirable features.

Figure 1 depicts the workflow of a typical MSI experiment. The generic MSI workflow is briefly discussed and challenges are identified. Some advances towards the solutions of these problems are pointed out and will be discussed in more detail further on.

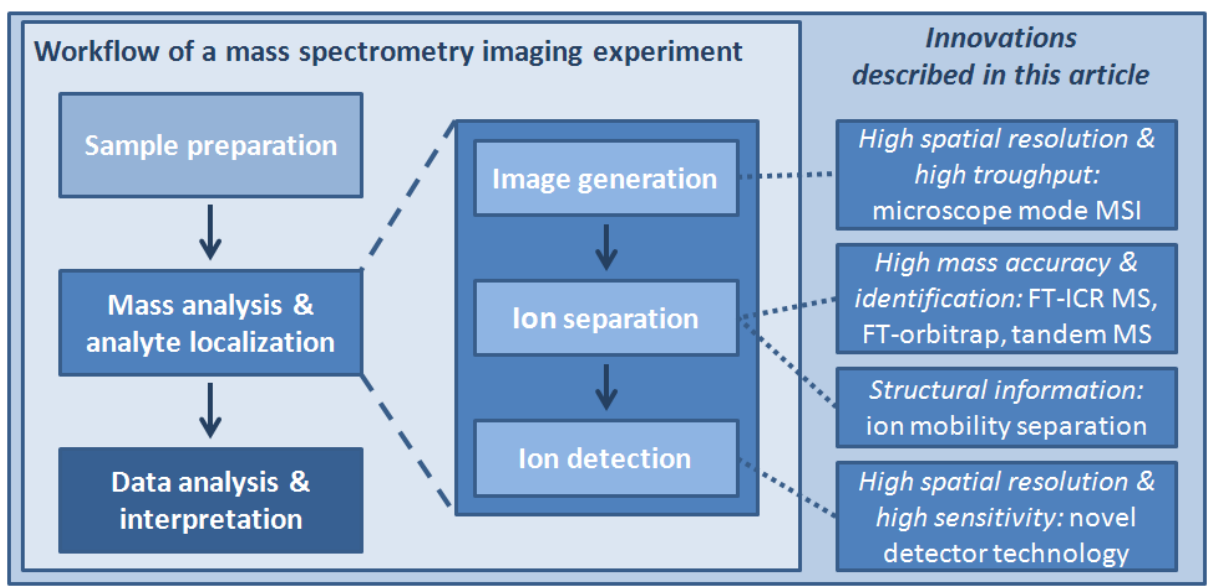

**Figure 1:** *General workflow of a mass spectrometry imaging experiment. Technological developments which will be highlighted in this article are indicated.*

Generally, MSI experiments consist three distinct parts, each of which poses its own challenges but also offers a variety of opportunities for specific sample analysis.

*(I) Sample preparation.* The first step in an MSI experiment is adequate sample preparation specifically targeted to the envisioned type of study. Sample preparation methods and protocols are an active area of research which merits in depth description and review [18-20] but is beyond the scope of this article.

*(II) Mass analysis & analyte localization.* The second step in the MSI workflow represents the mass analysis and the analyte localization, i.e. the actual mass spectrometry and imaging. Generally, MSI requires transferring solid state analytes into the gas phase. Then, the analytes are ionized. The different compounds are separated by their mass-to-charge ratio and eventually detected. In terms of MSI technology, this translates into three distinct areas of instrumentation: image generation, ion separation and ion detection.

*Image generation.* The image generation comprises the ionization and desorption as well as the formation of an ion optical image. The bio-molecules have to be desorbed and ionized such that analytes are transferred from the sample surface into the gas-phase. Among others [21-30], this is achieved with photons ((matrix-assisted) laser desorption ionization, (MA)LDI [21, 22, 31-35]) or primary ions (secondary ion mass spectrometry, SIMS [21, 22, 36-38]) in such a way that the spatial organization of the sample surface is retained. Different ionization methods call for specific sample preparation protocols and target different groups of analytes. The process of ion generation is a crucial part of MSI experiments [22]. The fundamentals of ion/ image generation and different ionization sources for MSI are not treated further detail in this review.

The ion/ image generation and the obtainable *spatial resolution* in MSI experiments are intricately related. In microprobe ("sample rastering") MSI, the spatial resolution is directly determined by the area of ion formation on the sample surface. Hence, most efforts approach the challenge of increasing the spatial resolution by decreasing the spot size of the ionization beam, i.e. the spot size of the laser or primary ion beam. Alternatively, the spatial resolution can be increased by stepping away from the sample rastering (microprobe mode) approach to microscope mode imaging [39]. This fundamentally different approach decouples the spatial resolution of the system from the desorption/ ionization spot by using magnifying ion optics and a position-sensitive detector. Microscope mode MSI is treated in detail in section 3 of this review.

The *analysis speed* and *sample throughput* of a MSI study greatly depend on the (frequency of) ion/ image generation. Microscope mode MSI experiments enable high sample throughput MSI studies due to the large desorption/ ionization spot as compared to microprobe mode MSI. In addition, sample throughput in microprobe mode MSI experiments has recently been targeted by several groups and will briefly be commented on in section 3.

By now, impressive technology is available for high throughput microprobe MSI studies on MALDI-time-of-flight (TOF) instruments. However, TOF-instruments intrinsically lack the mass resolving power and the mass accuracy that some bio-molecular and bio-medical imaging studies require Typically, ToF-MS systems have mass resolving powers below 40000 and mass accuracies after external calibration larger than 1-5 ppm which is insufficient for accurate compound identification. In the mass range $m/z$ < 500, accurate compound identification can be possible with TOF-MS. However, chemical noise complicates the compound identification in this mass range. To overcome these limitations, several other MS analyzers are under development to overcome the loss of chemical information due to instrumental mass resolving power and mass accuracy limitations (see ion separation/ mass analysis section).

*Mass analysis.* After desorption and ionization, the analytes are separated on the basis of their mass-to-charge ratio ($m/z$). The choice of mass analyzer depends on the required mass resolving power, mass accuracy, mass range and also availability. Generally, it is desirable and efficient to analyze all ions in one single analysis run.

*Fourier transform ion cyclotron resonance mass spectrometry* (FT-ICR-MS) determines the $m/z$ of an ion from its ion cyclotron resonance frequency in a static magnetic field. Similarly, the $m/z$ of an ion in a **FT-orbitrap** mass spectrometer is determined from its oscillation frequency in the electrostatic field of the trap. Since frequencies can be measured with a higher precision than a TOF, these instruments determine the ion mass with a significantly higher accuracy than a TOF-analyzer. Unlike TOF-MS measurements, the mass measurement in FT-MS is independent of the kinetic energy distribution of the ions. This is a clear advantage of FT-MS over TOF-MS where the spread in the kinetic energy distribution of the ions introduces mass inaccuracies in TOF-MS. MALDI FT-MS is the most prominent technique in revealing chemically complete and accurate information from a sample. This technique combines a high mass resolving power with a high mass accuracy. It resolves and identifies ions at a single nominal mass that are commonly present in MALDI imaging mass spectra [40-42] and provides a highly confident compound identification [43]. The accurate assignment of a molecular formula to a specific mass spectral peak is crucial to MSI studies. Direct molecular identification is only possible if an accurate mass measurement is available from a FT-ICR/ FT-Orbitrap MS measurement. Developments in this field will be treated in section 4 of this review.

***Time-of-flight –MS*** (TOF-MS) is a technique in which all ions are accelerated to a very narrow range of kinetic energies. Then, the ions are separated on the basis of their *m/z* in a flight tube. TOF-instruments intrinsically lack the mass resolving power and the mass accuracy that some bio-molecular and bio-medical imaging studies require. To overcome these limitations, several new techniques are under development to prevent the loss of chemical information due to limited mass resolving power and mass accuracy. If no accurate mass determination is possible, the molecular identity can be determined indirectly.

In particular, confident analyte identification from TOF-MS data can be achieved by **tandem MS**. Here, a selected parent ion is identified by (repeated) fragmentation and mass spectrometric analysis of the fragments [44]. While tandem MS capabilities are readily available for MALDI-instruments, they are still scarce for SIMS spectrometers where the primary ion beam readily damages the sample surface and ion yields are lower [45-49]. However, in recent years, cluster beam sources have greatly improved the sensitivity, the available mass range and the lateral resolution for biological SIMS [36, 50, 51]. The development of $C_{60}$-SIMS TOF-MS instruments with tandem-MS capabilities is a logical consequence [49, 52].

In addition, ***gas phase separation capabilities***, like ion mobility separation, have been incorporated. This technique separates ions of the same mass-to-charge ratio (but that constitute different bio-molecules) by their collisional cross section in a collision gas [53-55]. The gas phase separation step is typically conducted prior to the mass analysis in a TOF- spectrometer. Thereby, traveling-wave ion mobility separation reduces the complexity of the sample analysis. It also recovers errors in the spatial image introduced by unresolved peaks. Ion mobility separation will be treated in section 6 of this review.

***Ion detection.*** Rubakhin and co-workers describe that the ideal imaging technique for MSI can "simultaneously detect and identify multiple known and unknown compounds present in biological tissues with at least single-cell spatial resolution" [56]. Koppenaal and co-workers outline that the desirable analytical attributes of an ideal MS detector [57] are unity ion-detection efficiency, low or no noise, a high stability, the simultaneous detection of multiple ions, a wide mass-range and mass-independent response, a wide dynamic range, a fast response, a short recovery time and a high saturation level . They add operational attributes as a long life, low maintenance, easy replacement as well as a low replacement cost [57]. The spectrometer design or the envisioned application pose varying requirements to the MS(I) detector technology. For instance, experimental conditions may call for a detection system capable of handling high count rates (> $10^6$ counts/s) with minimal recovery time [57], detectors with a rapid readout and response (particularly in TOF-MS) [57], single or multiple ion detection [57], low noise (in the detector itself and the read-out electronics) for good limits of detection, high sensitivity, accuracy and precision [57]. Bio-molecular or bio-medical MS applications require detection uniformity in the mass range $10^2$ to $10^5$ *m/z* [57]. This poses a challenge to conventional secondary-electron MS detectors (like micro-channel plates (MCP) or photo-multiplier tubes) for which the mass response levels off for larger ions [57]. This results in a detection disadvantage for macromolecules of bio-molecular or bio-medical interest as for instance singly-charged, intact macromolecules like proteins studied by MALDI-MS [57]. This detection disadvantage for macromolecules calls for new detector technology. Novel, highly parallel detector technology for MSI is discussed in section 3.

***(III) Data analysis and interpretation.*** Finally, step three in the MSI workflow is the analysis and interpretation of the acquired data. MSI experiments return thousands of spectral channels at thousands of positions on the sample surface. Since mass spectra are measured at a raster of positions on the sample surface, the amount of data generated requires specific data treatment techniques. Therefore, software and algorithm development forms an integral part of the MSI field. Data analysis and interpretation are not extensively treated in this manuscript as a result of the chosen focus on emerging technologies. It needs little imagination that the bio-informatics aspects of MSI will need a substantial amount of work in the future. An algorithm which reconstructs the localization of specific analytes of bio-molecular or bio-medical samples in three dimensions and hence places the analyte localization in the context of tissue and/or organ functionality is presented in section 5.

In all of these steps, the selected method depends on the information that the experiment aims to reveal. In particular, the available mass spectrometry imaging technology provides different analytical capabilities and has its limitations. Appropriate instrumentation should be chosen to match the anticipated mass range, the mass resolution, the mass accuracy and the spatial resolution. Furthermore, the instrument's capabilities to perform additional analysis on the ionized sample, as for instance, fragmentation, tandem MS, spectroscopy or ion mobility separation should be considered. Each of the three steps in an MSI experiment has to be carefully planned and performed successfully for the generation of an adequate mass spectrum and corresponding molecular images.

**(3) Novel imaging and detection schemes for MSI: Developments to improve sample throughput, spatial image resolution and analyte detection.**

For many applications, TOF-MS mass analyzers are a good choice for both MALDI- and SIMS-MS. The speed, the obtainable sensitivity, the wide accessible mass range (up to *m/z* = 100k with MALDI) and the infrastructure for high sample throughput make TOF-MS the mass analyzer of choice for applications where extremely high mass accuracy and mass resolution are not imperative (or tandem MS is available). In this section, microscope mode MSI on a TOF-instrument is discussed as a high throughput, high spatial resolution MSI method.

In MSI, two acquisition modes can be distinguished namely the microprobe and the microscope mode (Figure 2).

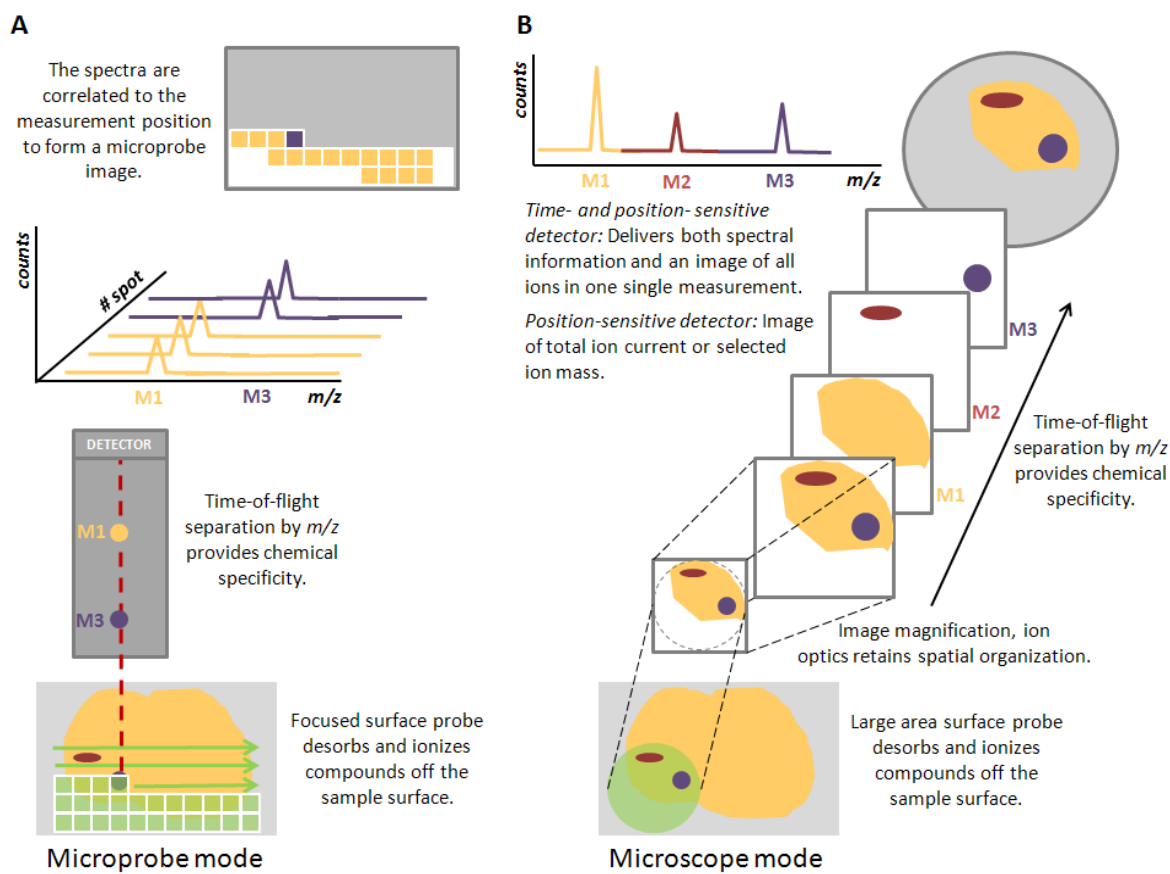

**Figure 2:** *A schematic representation of microprobe and microscope mode MSI (adapted from [39]). In microprobe mode MSI (A), the desorption/ ionization beam rasters the sample surface. At every raster spot, a mass spectrum is acquired. An image with a pixel resolution equivalent to the beam size is reconstructed after the experiment. In microscope mode MSI (B), a larger area desorption/ ionization beam illuminates the sample surface. Ion optics magnifies the molecular images and retains the spatial information. The molecular ion distributions are mapped on a position-sensitive detector.*

***Microprobe mode MSI.*** Most commonly, the conventional microprobe mode of acquisition is employed in MSI. In the microprobe mode, the sample is probed with a pulsed desorption/ionization beam (a laser beam in MALDI or a primary ion beam in SIMS), for which either the desorption/ ionization beam, the sample or both are moved with respect to one another. At every raster point, a full mass spectrum is generated. An image with a pixel resolution equivalent to the beam size is reconstructed after the experiment provided that the step-size/ accuracy of the sample or beam movement is not limiting. A disadvantage of small surface probe spots is the small number of ions generated for analysis. Using highly focused primary ion or laser desorption/ ionization beams, microprobe mode MSI has delivered high spatial resolution MSI studies. Pulsed primary ion beams [58, 59] and UV/ IR lasers in MALDI [60-63] have returned pixel sizes better than 4 μm and 7 μm, respectively. Specifically, a MALDI laser spot diameter of less than 1 μm and an according spatial resolution have been reported by the Spengler group [64]. However, a small surface probe area can bear the disadvantage of long measurement times and the fluence available for surface desorption and ionization may be limiting. Sweedler and co-workers have demonstrated MALDI microprobe MSI at a lateral resolution better than the laser spot size using over-sampling [65]. The Vickerman group has pioneered cluster SIMS-MSI, i.e. SIMS imaging with polyatomic primary ions [36, 49, 51]. Polyatomic ions generate less subsurface damage than atomic primary ions and hence enable imaging beyond the static limit (primary ion density results in surface damage of less than 1%).

*Microscope mode MSI.* Ion-microscope mass spectrometers provide an alternative technique to achieve high spatial resolution MSI [39]. Here, surface molecules are illuminated by a large area desorption/ ionization beam. In an ion microscope, the ionized analytes are extracted from the sample surface and imaged onto a position-sensitive detector. Initially all ions are accelerated by an electrical potential with the goal to give every ion the same initial kinetic energy. The instrument's ion optics typically achieves a rather narrow kinetic energy distribution such that negligible errors are introduced to the TOF-measurement and resulting mass assignment. The instrument's ion optics also magnify the initial ion distribution and retain the lateral spatial organization of the surface molecules on their flight path to the detector. Different molecular species are TOF-separated in the instrument's flight tube. The achievable spatial resolving power is on the order of 4 µm [39]. Importantly, the obtainable spatial resolution in microscope mode MSI is decoupled from the ionization spot size. Rather, the quality and the capabilities of the ion optics in combination with a position- (and time-) sensitive detector determine the lateral resolution (reference [66] shows a schematic representation of such a system). Very large sample surfaces are measured by few individual microscope spots which are then pasted together to form large microscope MS image [67]. The Heeren group has pioneered microscope mode MSI on a TRIple Focusing Time-of-flight (TRIFT) mass spectrometer (Physical Electronics, Inc., Chanhassen, USA) using a detection system based on microchannel plates (MCP) in combination with a charge-coupled device (CCD) camera [39], a delay-line-detector (DLD) [68] and active, pixelated detectors of the Medipix [69]/ Timepix [66] detector family.

*Throughput.* In ion microscopes, the desorption/ ionization beam size is typically between 200 µm and 300 µm, which is about a factor of 100 larger than a typical microprobe mode ionization spot. Comparing the microprobe and the microscope acquisition modes clearly indicates the high throughput capabilities of the latter. As an example, consider a microprobe pixel size (i.e. beam spot) of 2 µm × 2 µm (4 µm lateral resolution) and a sample area of 1 mm × 1 mm. Rastering the sample surface pixel by pixel will results in (1,000 µm × 1,000 µm) / (2 µm × 2 µm) = 250,000 measurement points. Such a measurement can last several hours. In microscope mode MSI, a field of view of 200 µm × 200 µm is available. So, for the same sample size of 1 mm × 1 mm, only (1,000 µm × 1,000 µm) / (200 µm × 200 µm) = 25 measurement points are required. A lateral resolution on the order of 4 µm is readily achieved with microscope mode MSI using an appropriate ion optical magnification factor and position-sensitive detector. In microscope mode MSI, the large ionization spot can result in large ion loads on the detection system. Several position-sensitive detection systems saturate at high ion loads such that the ion generation has to be reduced and the integration time per raster position has to be increased to ensure adequate ion imaging. This can be compensated by a combination of high desorption/ ionization beam repetition rates and detector technology that can accommodate high ion loads. Hence, ion microscopes enable fast, automated, high resolution large area imaging provided that adequate, i.e. fast and position- (and possibly time-) sensitive, detector systems are employed to record high quality molecular images [70].

The improvement of sample throughput in microprobe mode MSI has recently been studied by few groups. McDonnell and co-workers have designed and commissioned an automatic sample loading system for time-efficient, round-the-clock spectrometer operation [71]. An alternative approach to high throughput measurements is an increase in the raster frequency, which results in a decrease in the measurement time without sacrificing analytical information. The Caprioli group has presented such an approach, in which they reduce the measurement time by up to a factor of two by operating a MALDI-TOF MSI system in a continuous scanning mode at a comparatively high laser repetition rate of about 5 kHz [72]. Stoeckli and co-workers generate images of drug distributions in rat sections within less than 15 minutes by using a 1kHz laser repetition rate and a rastering speed of about 18 mm/s [73]. Sample throughput is a key factor in distinguishing MSI experiments as a robust, analytical technique for bio-molecular and bio-medical research.

***Detector Technology.*** The first microscope mode MSI implementation projected the ions on a position-sensitive detector assembly which consisted of an MCP followed by a phosphor screen and a charge coupled device (CCD) camera [39]. This detector scheme allowed for the proof-of-principle of the technique. However, the MCP, phosphor screen, CCD camera assembly suffered from the limitation that it cannot link the ion TOF (encoding the *m/z*) and the spatial distribution. During one acquisition, the total ion current can be measured without *m/z*-specific localization on the sample. Or alternatively, a particular *m/z* species can be selected to pass to the detector by a pair of fast-switching electrodes, called the electrostatic blanker. The combination of several mass-selected images from separate measurements then gives information on the sample composition and according spatial distribution. This detector approach calls for time-consuming, highly repetitive measurements on quickly depleting biological samples.

A second, more advanced implementation of microscope mode MSI used a delay-line anode for ion detection [68, 74]. A delay-line detector can measure both the ion arrival position and time precisely [75] and thereby lifts the limitations of mass-selected imaging. In addition, the ion arrival time can be measured precisely by registration of the TOF signal from the MCP detector and correlating this information with the position information from the delay-line anode [76]. A drawback of delay-line detectors is that it can only register a small number of ion events that simultaneously arrive at the detector [77]. This renders it unsuitable for high count rate (MALDI-) MSI experiments. Furthermore, the mass-resolved image reconstruction can be time-consuming. Ion optical tuning of the instrument is tedious due to delayed image feedback.

Recently, the implementation of an in-vacuum pixel detector for position- [69, 78] and time-resolved [79-81] electron and ion imaging was demonstrated and applied to MSI on an ion microscope [69, 81]. This novel detection approach uses a fully-integrated, active pixel detector of the Medipix/Timepix detector family [82-84]. Such a detector assembly consists of a chevron MCP followed by four bare Timepix chips (with 256 × 256 pixels of 55 μm × 55 μm each per chip). Each of the Timepix pixels can return (1) the position of impact of an ion via its pixel address and (2) the ion time-of-flight with respect to an external trigger with a maximum time resolution of 10 ns. The maximum TOF that can be measured per pixel is determined by the counter depth (counter overflows at 11810). At the maximum time resolution of 10 ns, the counters can measure maximum ion TOFs of about 118 μs. Conveniently, the simultaneous acquisition of the ion TOF and arrival position removes the need for mass-selection using an electrostatic blanker. Hence, mass-resolved images are generated during a single imaging acquisition. Unlike the delay-line detector assembly, this pixelated detector facilitates parallel detection of multiple molecular species and can accommodate significantly more ion hits simultaneously. This bears great potential for both higher count rate and large area high resolution molecular imaging experiments. Indeed, peptide and protein mass spectra have been generated over mass range up to 80 kDa [81] on an ion microscope. An in depth comparison of the quality of the Timepix acquired TOF spectra to established detection techniques such as the recording of mass spectra using a time-to-digital converter (TDC) and an analog-to-digital converter (ADC) was performed [81]. This novel detector approach outperforms the conventional technology easily in the detector dynamic range, the signal-to-noise ratio, the multiplexed detection on several detector elements, the available mass range, the detection homogeneity, the ability to detect single particles, the ability to resolve isotopes and the detector response to varying MCP gains [81]. Note that the current performance of this detection system is still limited by few technological shortcomings. A future pixelated chip for microscope mode MSI should incorporate multi-stop pixels (i.e. pixels which can accommodate more than one hit per measurements cycles), a larger counter depth (i.e. longer TOFs can be measured) and TDC bins on the order of 100 ps such that the achievable mass resolution becomes comparable to present commercial MSI instrumentation. An ion microscope equipped with such a detector assembly will provide a very powerful MSI instrument for bio-molecular or bio-medical studies.

***Considerations on the spatial resolution.*** On an ion microscope, the ultimate spatial resolution is determined by the ion optical design and its ion optical aberrations. Still, ion optical aberrations often do not turn out to be the limiting factor to the spatial resolution of the system. Rather, the accessible spatial resolution is determined by the resolution of the position-sensitive detection system. Both the MCP, phosphor screen, CCD camera assembly [39] and the MCP, delay-line detection system [68, 74] do not deliver spatial resolutions on the order of the ion optical aberrations. For example, the Heeren group ion microscope (used in [39, 68, 69, 74, 81]) delivers a maximum ion optical magnification of 100x. The spatial resolution of this system is limited by ion optical aberrations beyond 600 nm [85]. This means that at the maximum ion optical magnification, the detector spatial resolution should be no less than 60 µm (= 100 × 600 nm). In this scenario, the 60 µm on the imaging detector probe 600 nm on the sample surface. The detection system is not limiting the spatial image resolution. The pixelated Timepix detection system consists of an array of 55 µm × 55 µm pixels. Hence, the full resolution delivered by the ion microscope can be exploited with this detection system.

***Pushing detection limitations.*** As mentioned before (section 2), the ideal imaging technique for MSI can "simultaneously detect and identify multiple known and unknown compounds present in biological tissues with at least single-cell spatial resolution" [56]. An evaluation of the three afore mentioned detection systems (Table 1) against these criteria, indicates that the MCP, phosphor screen, CCD camera system and the MCP, delay-line detections system have difficulties meeting all those criteria. It should be noted that individual characteristics (points of comparison below) of those systems may be significantly better in detection systems tuned to specific needs of a particular experiment. An example is the extremely high timing resolution of about 18 ps reported for an MCP, delay-line detector system by Vredenborg and co-workers [76]. However, only a very limited number of simultaneous hits can be accommodated by the detection system. To our knowledge, no MCP, phosphor screen, CCD camera system or MCP, delay-line detections system has been reported that combines all of the exceptional performance aspects which could turn those systems into candidates for the "ideal" MSI detection system. The MCP, active pixel detector assembly comes significantly closer to the desirable capabilities of an "ideal" detector assembly for MSI. In particular, bio-molecular or bio-medical MS applications can benefit from the uniform ion detection, i.e. from the minimized detector roll-off due to sub-saturation MCP operation, in the mass range up to 80 kDa.

Table 1: Comparison of three different detection systems used in microscope mode MSI to the "ideal" MSI detection system.

| Point of comparison | Ideal MSI detection system [56, 57] | MCP + phosphor screen + CCD camera [39] | MCP + delay-line anode [68, 74] | MCP + active pixel detector (Timepix) [81] |
|---|---|---|---|---|
| Position- and time- sensitive detection | Yes. | No. | Yes. | Yes. |
| Single cell spatial resolution (± 10 μm) | Yes. | Yes.* | Yes.* | Yes.* |
| Unit ion-detection efficiency | Yes. | Yes.† | Yes.† | Yes.† |
| Simultaneous detection of multiple ions. | Yes. | Yes.† | Yes.† | Yes.† |
| Count rates | Yes, > $10^6$ counts/s. | Yes./No.$ | No.** | Yes, > $10^6$ counts/s. |
| Noise | Low or no noise. | Electronic noise. | Electronic noise. | No electronic noise. |
| Stability | High. | High. | High. | High. |
| Mass range‡ | Wide. | MCP high mass roll-off limited. | MCP high-mass roll-off limited. | ≤ 80 kDa, MCP not in saturation. |
| Mass-independent response | Yes. | No, high mass role-off due to MCP. | No, high mass role-off due to MCP. | Reasonable, MCP not in saturation.** |
| Dynamic range | Wide. | 1 | 1 | 262,144 detection elements for parallel detection |
| Response | Fast. | Phosphor decay time on order of 100s of μs. 138 ps (TDC) | 25 ps (TDC) | 10 ns (pixel clock) |
| Recovery time | Short. | Read-out time dependent. | Read-out time dependent. | Read-out time dependent. |
| Saturation level | High. | 1 | 1 | 262,144 detection elements for parallel detection |
| Read-out/ dead time | Rapid/short | ± 1 ms | < 5ns | 1 ms |
| High (spectral) sensitivity | High (++). | ++ | + | ++ |
| Timing precision | High. | 138 ps (TDC) | 25 ps (TDC) | 10 ns (pixel clock) |
| Mass accuracy+ | High. | Determined by clock stability. | Determined by clock stability. | Determined by clock stability. |
| Spatial accuracy | High. | Determined by | Determined by | Determined by |

|   |   |   | ion optics. | ion optics. | ion optics. |
|---|---|---|---|---|---|
| Spatial precision[‡],[††] | High. | 4 µm | 4 µm | 3 -4 µm |
| Life time | Long. | MCP limited. | MCP limited. | MCP limited, but prolonged by sub-saturation operation. |
| Maintenance | Low. | Low. | Low. | Low. |
| Replacement | Easy. | Standard assembly, readily exchanged. | Involved, bulky assembly. Major effort to replace. | Compact assembly, readily changed. |
| Replacement cost | Low ($). | $ | $$$ | $ |

[*] On an ion microscope imaging mass spectrometer with an appropriate magnification factor.

[†] Yes, at sufficiently low count rates. Note, that the active area of a MCP is 70% - 80%. Hence, any detection system in combination with a MCP cannot deliver unit quantum efficiency. However, it is advantageous if the second detector stage is capable of registering the MCP electron shower with (near) unit detection efficiency.

[**] Approximately, 10 counts/trigger have been realized at repetition rates of about 2 kHz [68].

[‡] On ion microscope imaging mass spectrometer [85].

[**] Individual Timepix pixels are single-stop TDCs. Ions generated on the same position on the sample surface and hence impinging on the same pixels on the detector cannot be distinguished in a single shot experiment. There is a detection advantage for the lower mass ion. Practically, this effect is not limiting since experiments are typically based on multiple laser shots and the effect is leveled out by sufficient measurement statistics [81].

[††] The spatial precision on the sample surface is evaluated in terms of the spatial resolving power, i.e. the sharpness of the sample feature edges. On an ion microscope, the spatial precision depends on the ion optical magnification factor and the element/ pixel size on the imaging detector. In microprobe measurements, the spatial precision would be determined by the combination of the ionization spot size and the element size of the detection system.

[+] The dependence of the mass accuracy on the (stability of the) detection system is evaluated. Note, that generally the mass accuracy is dependent on an instrument function that takes into account the type and the mass accuracy of the mass analyzer itself, the detection system and also the sample surface homogeneity.

[$] The phosphor screen can in principle deal with high count rates when operated in a signal integration mode, i.e. when no spatial information is collected from the phosphor screen. Under high count rates conditions, the light intensity over-exposes the CCD-camera and all spatial information is lost, making it unsuitable for direct imaging MS experiments under these conditions.

***Active pixel detectors for MSI.*** Considering the high potential of active pixel detectors for MSI, the Medipix/ Timepix detector –as a representative of this novel type of detection system- and its outstanding capabilities are outlined in more detail. Very few active pixel detectors with comparable functionality are currently available. An interesting alternative chip is currently being developed by the NA62 collaboration at CERN[86-90]. This read-out chip achieves a timing resolution of about 100 ps and is designed to accommodate about 73k hits per second per pixel. However, the 45 × 40 (= 1800) pixel matrix with a pixel size of 300 µm × 300 µm compromises the desirable spatial resolution and covers only a rather small area. In addition, the power consumption of the chip is about two orders of magnitude higher than for the Timepix chip, which renders practical, in-vacuum implementations on mass spectrometers difficult.

The Medipix/ Timepix detector family is developed within the Medipix collaboration hosted by CERN [91]. The chips of the Medipix detector family in combination with a detection medium -often a semiconductor like silicon bump-bonded on top- belong to the class of hybrid pixel detectors. There are two distinct types of chips within the Medipix detector family: the Medipix single photon counting chips and the Timepix chips, which in addition to the single photon counting capabilities can also be set to measure the time-of-arrival of an event with respect to an external reference signal or to determine the amount of charge deposited per pixel.

Semiconductor materials bump-bonded to Medipix2 chips are silicon, gallium-arsenide, cadmium telluride, cadmium-zinc-telluride or germanium depending on the application of the detection system. Typically, a 300 µm silicon sensor (slightly n-doped high-resistivity silicon with a p-type implementation in every pixel) is bump-bonded on top of a Medipix chip. On the entrance side, the sensor layer is coated with an aluminum layer of about 150 nm. Through this Ohmic contact, the sensor material is biased by applying a voltage of about 100 V across the sensor. Depending on the polarity of the applied bias voltage, an electron or a hole current can be collected by the pixels. In silicon, every 3.6 eV of deposited energy creates one electron-hole pair. Hence, the amount of charges generated in the sensor material is directly proportional to the energy deposited by the impinging particle. With such a sensor layer photons and electrons can efficiently be detected provided that the photon or electron kinetic energy exceeds the detection threshold of about 4-5 keV.

When used for X-ray and electron detection, the detection medium converts incident particles into electron-hole pairs, which are collected in the charge-sensitive amplifier of the CMOS (complementary metal oxide semiconductor) read-out chip. Ions will usually not be accelerated to sufficient energies to penetrate into the sensor layer. However, ions can be detected indirectly by placement of an MCP in front of the detector [69, 92]. The Medipix detector then registers the electron shower produced by each ion impact on the MCP. The particle counting properties consist of the ability to count events that generate a number of electron-hole pairs within a user-defined threshold/ energy window.

The threshold energies that are chosen for the discriminator levels lie well above the noise levels of the pixels (on the order of 150 electrons). Therefore, electronics noise free measurements are possible, while background noise of chemical origin or due to (cosmic) radiation can still be picked up. Three additional adjustment bits can be used to equalize the pixel-to-pixel response over the full pixel array.

Measurements involving time-of-flight or ultra-high resolution measurements are performed using Timepix chips. The Timepix chip [84] is derived from the Medipix2 chip design. The dimensions and geometry of the chip are identical to its predecessor but the functionality on the pixel level is different. Each pixel can be individually selected to operate in one of three modes:

(1) the counting mode, in which each pixel counts the number of events;

(2) the time-of-flight (TOF) mode, in which the occurrence time of an event is measured with respect to an external trigger/shutter signal;

(3) the time-over-threshold (TOT) mode, in which the time is measured during which the charge resulting from the event exceeds the detection threshold level.

The maximum measurement time in TOF and TOT mode is determined by the pixel counter depth in combination with the measurement clock speed. The Timepix pixel counter is a 13-bit pseudo-random counter. The maximum counter value of the Timepix chip is 11810, i.e. 11810 is the pixel overflow value. Therefore, at the maximum clock speed of 100 MHz (i.e. 10 ns clock cycles), a maximum measurement interval of 11810· 10 ns = 118.1 µs is available.

In the implementation on the MSI ion microscope [81], it was chosen to use chips without a sensor layer, so-called bare chips, to improve the response to electron showers through reduced in-sensor electron diffusion.

**(4) Identification strategies: Imaging at high mass resolving power and high mass accuracy**

One of the key challenges in mass spectrometry imaging is the ability to accurately identify the molecular species imaged. Until recently innovations were predominantly focused on new methods of image generation with time-of-flight MS. TOF based MS methods are shown to offer high spatial resolution and throughput but lack the mass resolving power and mass accuracy needed for the identification of observed peaks. Unraveling the complexity of molecular profiles at biological surfaces is however hampered by the limited mass resolution, sensitivity, dynamic range and spatial resolution of conventional TOF based mass spectrometric systems. The lack of mass resolving power can result in obscured spatial details when two closely neighboring peaks have different spatial distributions and are not resolved in the mass spectral domain. The availability of high resolution and high mass accuracy mass spectrometers such as FTICR-MS [93] and FT-Orbitrap-MS [94] for imaging MS is providing the needed capabilities for accurate molecular identification.

The Orbitrap mass analyzer [94, 95] is an electrostatic trap wherein tangentially injected ions rotate around a central electrode, being confined by applying an appropriate voltage between the outer and central electrodes. Mass analysis is based on image current detection of frequencies of axial oscillations. Therefore, its extent of mass accuracy is limited by the same factors as FTICR. The introduction of the LTQ-Orbitrap has revolutionized high performance, high throughput mass spectrometry. It is rapidly making its way into high performance MS imaging. The first example that demonstrated the usefulness of high mass resolving power in an MS imaging strategy was provided by Taban and co-workers [41] on an FTICR-MS. Analyzing endogenous peptide distributions on a rat brain, they revealed new spatial detail by separating mass spectral features that were unresolved with comparable lower resolution experiments. Figure 3 shows how the increased mass resolving power of an FT-ICR-MS system reveals such new spatial detail. Recently, the spatial resolution of FT-Orbitrap-based MSI was improved in an atmospheric pressure scanning microprobe matrix-assisted laser desorption/ionization mass spectrometry (AP-SMALDI-MS) experiment using a tightly focused laser beam [96]. Verhaert and co-workers [97] have employed high mass accuracy MSI procedures with an FT-Orbitrap to localize physiologically active peptides in neuronal tissue from American cockroach (Periplaneta americana) neurosecretory organs. Their results clearly illustrate that high mass accuracy and high mass resolving power of the Orbitrap analyzer are now routinely achievable in direct tissue analysis and molecular imaging experiments. This high mass resolution MS imaging approach now allows the direct identification and structural analysis of lipids, peptides and proteins from a variety of complex biological surfaces. This detailed information provides new fundamental insight in dynamic biological processes such as cell and tissue differentiation and signaling.

Fourier transform based technologies for imaging bring forward new challenges. The Fourier based technologies provide exquisite molecular detail but are inherently slow. The high mass resolving power requires long transients to be acquired for each pixel. Transients of several seconds per pixel have been reported. This limits the applicability of FTMS based tissue imaging to compounds that are stable to in-vacuum degradation. Alternatively, pressure high-spatal resolution MALDI could be employed [96] to address the in-vacuum degradation issue. New mathematical techniques such as the application of a filtered diagonalization method (FDM) [98] as an alternative for the Fourier transform could be instrumental in improving total acquisition time. The total data volume is a significant limitation on the total analysis time in the generation of high resolution molecular images. This applies for both high spatial resolution as well as high mass resolution approaches. Mass spectrometry imaging at higher resolution therefore also requires developments in data storage, processing and visualization methods. A significant speed up in statistical analysis was recently introduced by the application of graphical processing units (GPU's) for principal component analysis [99]. Parallel preprocessing with cluster approaches demonstrated for high resolution LC-MS datasets [100] can now readily be applied for rapid high resolution MSI. It is clear that on many levels the high mass resolution approaches provide detailed molecular information that complements the MS based molecular imaging technologies. Consequently, more and more researchers are resorting to multi-modal imaging approaches in which as much molecular information is gathered from one single piece of tissue.

Proteomics strategies have benefited from the availability of the high mass accuracy methods for a substantial time already. It has resulted in improved throughput and identification capabilities of a wide variety of peptides and proteins and their post-translational modifications [101]. Recently, these methods are established as proteomics based MS imaging workflows [4]. In the previous section we argued that increased mass resolving power reveals more spatial and molecular detail. Increased mass accuracy will, similar to the developments in proteomics, enhance the identification capabilities for imaging MS experiments. Imaging small molecules (< 400 *m/z*) benefits directly as high accuracy allows the direct determination of the elemental composition of a compound. This capability is particularly useful for drug and metabolite imaging [102].

The increased mass accuracy also improves protein identification capabilities in the analysis of tissue section where an on-tissue digestion protocol has been applied. This proteomics imaging strategy employs a similar protein identification workflow as in conventional LC-tandem MS applied to each pixel. The identification is clearly more challenging due to the lack of chromatographic separation techniques after local on-tissue digestion. Here, the combination of high mass resolution (separation power) and high mass accuracy (identification power) is crucial. The high mass accuracy improves the peptide identification scores in conventional protein database searches. Peptide Mass Fingerprinting (PMF) and tandem MS based protocols are commonly used in concert.

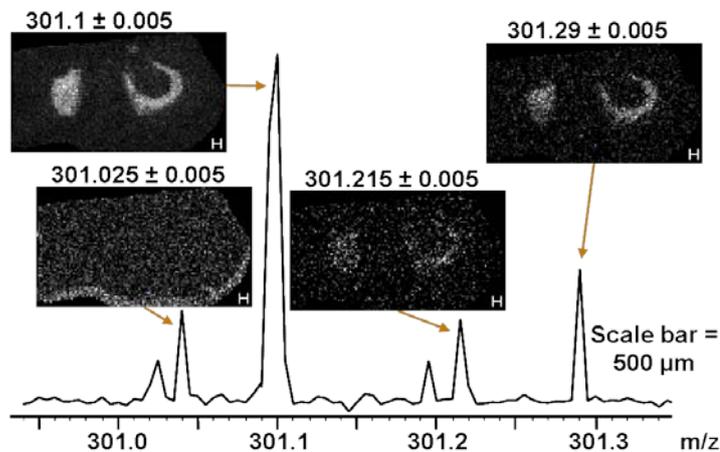

**Figure 3:** *High mass resolving power SIMS-FTICR imaging study of a mouse brain. This figure illuistrates how within 0.3 Dalton several molecular peaks can be resolved that all have different spatial distributions. (Previously unpublished result, courtesy of D.F. Smith, R.M.A. Heeren and L. Pasa-Tolic).*

Naturally, the combination of both high mass accuracy measurements and very high spatial resolving powers is desirable to meet the demand for detail in the image and in the spectral dimension. Spengler and co-workers used a coaxial objective to generate a laser focus smaller than 5 μm on several Thermo FT-MS systems (Thermo Scientific GmbH, Bremen, Germany). This approach was employed to generate various high resolution MALDI based molecular images at the cellular level [103]. Smith and co-workers developed a MSI instrument which combines the high spatial resolution capabilities of a Buckminsterfullerene ($C_{60}$) cluster ion source (Secondary Ion Mass Spectrometry) with the high mass resolving power and high mass accuracy of FTICR MS [104]. This novel instrument features a commercial $C_{60}$ primary ion source, 12 T solariX FT-ICR mass spectrometer and tandem MS capabilities (Bruker, Billerica, USA). Smith and co-workers demonstrate a microprobe MSI measurement on mouse brain at a 40 μm pixel size, a mass resolving power better than 100,000 ($m/\Delta m_{50\%}$) and a mass accuracy well below 1 ppm. Extremely high spatial resolution approaches are ultimately limited by the efficiency of the ionization method. How many molecules are available for ionization at the ultimate spatial resolution and how many of those molecules will be ionized, i.e. what is the ionization efficiency? If the ionization efficiency is less than 100%, which holds for most ionization methods, the ionization efficiency will determine a practical achievable resolution. One exception is the microscope mode discussed earlier in this review (section 3).

The previous section has emphasized mass accuracy and mass resolving power as an essential element of MSI identification strategies. One other crucial element in the identification of bio-molecules at surfaces in MSI experiments is tandem mass spectrometry. In particular in the absence of sufficient mass accuracy, it is imperative to perform structural analysis by means of selective dissociation to establish or confirm the identity of the species imaged. The preferred dissociation technique in MS imaging studies is collision induced dissociation as it is fast, easy to control and the fragment spectra are reasonably well understood or interpretable. A number of tandem MS imaging studies is focused on tryptic peptide identification for proteome imaging studies. These studies can be complicated by overlapping isobaric ions. Ion mobility separation, discussed later in this review (section 6), prior to tandem MS has proven useful to enhance peptide identification scores from tandem MS studies.

Tandem MS is significantly limited by the signal persistence in most MS imaging experiments. As a result it is not possible to perform tandem MS analysis of each and every peak found in a pixel. Intelligent tandem MS target selection is becoming a prerequisite for focused molecular imaging studies. Target lists based on prior quantitative proteomics analysis from tissue homogenates is a strategy followed by a number of researchers in proteomics. This allows the utilization of a priori knowledge on interesting up- and down-regulated proteins relevant for the experimental condition that is tested. Alternatively, a regular MS imaging experiment can be used to identify interesting spatial molecular structures that can subsequently be utilized for further tandem MS imaging experiments. Targeted tandem MS profiling studies on areas of interests can also be employed to identify the molecules found in those areas.

With the introduction of Multiple Reaction Monitoring (MRM) based imaging strategies [105] it is now also possible to perform quantitative tandem MS studies. Pirman and Yost [106] have used a similar strategy to quantitatively image the distribution of acetyl-L-carnitine in mouse brain tissue sections.

In summary, it is imperative that imaging MS researchers use a smart combination of high mass resolution, high mass accuracy and complementary tandem MS strategies to identify molecular distributions found on biomedical tissue surfaces.

**(5) Adding an new dimension: 3D-MSI**

Anatomical atlases based on optical images are widely used for anatomical and physiological reference in the medical profession. These atlases are employed to evaluate healthy and diseased tissue for diagnostic and treatment purposes. Often general disease treatment or management protocols are established solely based on morphological aberrations observed in biopsies. A new method needs to be established that provides a molecular basis for these anatomical atlases. This requires a molecular imaging method that provides a detailed insight in the spatial distribution of a broad range of elements and molecules. These molecular tissue atlases should combine 3-dimensional position information with molecular information.

Mass spectrometry imaging is typically a discipline in two spatial dimensions (2D). By positioning the laser or the primary ion beam to different locations on the surface results in a comprehensive set of mass spectra. A two-dimensional image of each individual mass spectral peak (*m/z*) can be generated. The 2D spatial information in combination with the *m/z* information of the spectra is often referred to as a datacube. The generation of a three-dimensional (3D) dataset requires an additional z-dimension. In MALDI-MSI this is achieved by successive tissue sectioning with well defined and measured spatial intervals [107, 108]. MSI-data is acquired and subsequently processed to reveal the three-dimensional molecular features. Figure 4 shows an example of a 3D MALDI-MSI approach in which lipid volumes in a xenografted tumor are revealed. The processing protocol can include spectral and spatial binning to reduce the total dataset size prior to molecular feature visualization. Different software tools have been developed to visualize these three dimensional reconstructions [109, 110]. Individual mass spectral peaks can be simultaneously displayed using a color scheme in which each color represents a specific molecular feature.

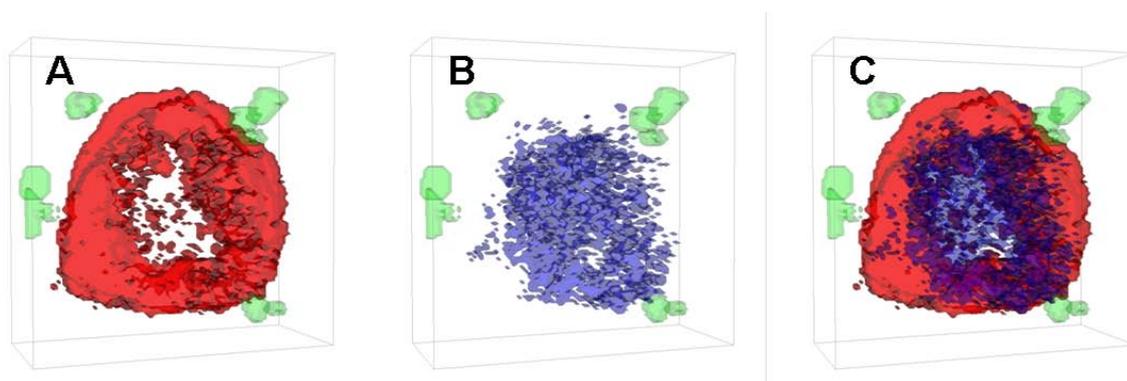

**Figure 4:** *Three dimensional molecular reconstruction of a xenografted breast tumor using fiducial marker alignment of individual sections [3]. This reconstruction is generated from molecular ion distributions obtained from the central section of the breast tumor. The fiducial marker ion Chresyl Violet at m/z 262.1 is shown in green in all images. a) Distribution of the molecular ion PC(16:0/0:0) [M+H]+ at m/z 496.3 is shown in red. b) Distribution of molecular ion PC(18:1/18:1) [M+K]+ at m/z 824.5 is shown in blue. c) The overlay of (a) and (b). Figure courtesy of K. Chughtai.*

An alternate approach towards the generation of 3D imaging MS datasets at cellular resolution is found in the SIMS domain. The introduction of ion-cluster based sputter guns for depth profiling allowed gentle removal of surface layers without the induction of subsurface damage. Sputter sources that employ $C_{60}$ molecular ions or large argon cluster ions [111, 112] can sputter with a depth resolution of tens of nanometers. This depth resolution enables 3D MSI on a subcellular scale with dynamic SIMS. Chandra and co-workers demonstrated the generation of 3D elemental images of the mitotic spindle from T98G human glioblastoma tumor cells with dynamic SIMS [113]. This study demonstrated that 3D SIMS imaging was essential for the analysis of mitotic cells, where specialized regions such as the mitotic spindle were hidden beneath the cell surface. A combination of sputter and analysis cycles with these sources is now employed for SIMS based molecular 3D imaging. This was demonstrated by Fletcher et al in their visualization of the 3D distribution of phosphocholine and inorganic ions in single cells using a TOF-SIMS approach [114].

In the future, the generation of 3D molecular atlases can be employed to populate comprehensive molecular tissue bio-banks, i.e. biological databases, for different tissue types. These atlases can be used to study local molecular changes involved in disease and therapy. Using e-science, grid or cloud technology these immense data volumes can be validated and made available to the biomedical research community for comparative studies. Tissue bio-banks should be complemented with patient data, comprehensive proteomics data of tissue homogenates, microarray data and conventional histological images. This has the potential to generate new insights in diagnosis, development, treatment and prognosis of disease.

**(6) Ion mobility: adding the shape of a molecule**

Alternate methods that are able to unravel the complexity of biochemical surfaces have become available to researchers in mass spectrometry imaging. It has already been established that cells, a tissue section, a tissue extract or a body fluid contain a huge amount of chemical and biological information. A single analytical method is usually insufficient when a better understanding of the function of these different biological systems is targeted. It has been illustrated how high mass resolution techniques can deal with this complexity. An alternate method is to add a gas phase separation technique to the MS imaging workflow. Hyphenated mass spectrometry techniques (e.g. LC-MS, GC-MS) are common in the analytical domain and are widely used for increased sensitivity and selectivity for bio-chemical analysis. Ion mobility spectrometry (IMS) combined with mass spectrometry imaging is one of these approaches that has the potential to provide direct insight in the shape, structure and position of bio-molecules. The combination of mass spectrometry with ion mobility spectrometry has already proven an extremely successful technique for determining the structures of ions in gas phase as it allows the separation of different structural isomers [115, 116]. The addition of a molecular imaging component enhances the molecular detail provided [55].

MALDI-ion mobility MS brings an added value to MALDI-MS tissue imaging by the separation of different compounds families such as lipids and proteins [54, 117]. The ion mobility cell, positioned between a quadrupole and a time-of-flight analyzer, allows the separation of structural isomers, or compounds with similar *m/z*, which cannot be separated by *m/z* only (due to TOF-instrument limitations). Different ion conformations have different collisional cross-sections and result in different drift times. This particular property is comparable to liquid chromatography separation and allows the separation and identification of chemical families such as matrix, drugs, lipids and single or double charged peptide ions by their retention time inside the ion mobility cell. Stauber and co-workers [5] demonstrated the ability to separate isobaric ions with ion mobility prior to tandem MS in a MALDI imaging experiment. The fragmentation profiles of different ions enabled a database search with increased specificity by pooling fragments associated to only one parent. The comparison between Mascot scores [118] with and without drift time fragment pooling shows the power of the added IMS separation. Gas phase ion mobility separation is frequently used to separate small matrix ions and their clusters from tryptic peptides and lipids prior to tandem MS. However, signal depletion prohibits extensive tandem MS imaging experiments. This requires an efficient selection of relevant precursor ions that can be realized with ion mobility separation. For example only tryptic peptide ions can be selected for fragmentation when protein identification is targeted in an imaging experiment [119].

The SYNAPT HDMS (Waters Corporation, Milford, MA) is one of the widely implemented ion mobility instruments with MSI capabilities. This instrument has a quadrupole orthogonal acceleration TOF geometry and is equipped with a travelling wave (the so-called T-wave) ion mobility device located between the quadrupole and the TOF analyzer. The ion mobility separator used consists of three consecutive traveling wave regions. The first traveling wave (trap) is used to store ions when IMS is performed, to maximize the duty cycle of the IMS. The next travelling wave section is the actual ion mobility separation device. The final traveling wave device (transfer) is used to transfer ions from the ion mobility separator to the TOF mass analyzer maintaining the ions' separation. Collision induced dissociation (CID) can be achieved in either the trap or transfer T-wave or in both. The instrument is equipped with an interchangeable MALDI source, which can be replaced with atmospheric ionization (AI) sources, such as electrospray ionization (ESI) or atmospheric pressure chemical ionization (APCI). This configuration allows for a variety of atmospheric and in-vacuum MSI experiments. MALDI is performed in an intermediate-pressure environment ($9 \times 10^{-2}$ mbar) using a frequency-tripled Nd:YAG laser (355 nm). Imaging data are obtained in the microprobe mode with a typical resolution of 80 μm. The time per pixel, pertaining to the throughput for MS imaging, is similar to that in conventional TOF-based imaging experiment. The increased dynamic range stems from the fact that per laser shot multiple orthogonal TOF-spectra are acquired following ion mobility separation.

MALDI-IMS-MSI has the ability to improve the imaging of some drugs, metabolites, lipids and peptides by separating such ions from endogenous or matrix-related isobaric ions. It has been applied in variety of studies ranging from whole body imaging of drug dosed rats to imaging signaling proteins in oncological studies [15, 119].

**(7) Summary**

The ideal or perfect mass spectrometry imaging experiment on tissue can still not be performed as a result of several limitations in spatial resolution, molecular identification capabilities and speed/throughput. In this review, we have shown several technological innovations that address these hurdles towards the perfect mass microscope. All of the innovations improve a specific aspect of the MSI workflow described in Figure 1. As such, the insight in the molecular organization of molecules on tissue surfaces is drastically improved. Having stated this, it is also clear that it is very difficult to integrate all of the innovations discussed in one single instrument. This implies an intrinsic need for multi-modal imaging experiments, in which different molecular detail is brought together. This in turn defines a need for more and improved bio-informatics tools that can integrate and validate this multi-modal data. This will, in our opinion, be one of the major areas of innovation for the years to come.

**Acknowledgements**

Part of this research is supported by the Dutch Technology Foundation STW, which is the applied science division of NWO, and the Technology Programme of the Ministry of Economic Affairs, Project OTP 11956. This work is part of the research program of the "Stichting voor Fundamenteel Onderzoek der Materie (FOM)", which is financially supported by the "Nederlandse organisatie voor Wetenschappelijk Onderzoek (NWO)".